\documentclass{emulateapj}

\bibstyle{aas}

\begin{document}
\bibstyle{aas}

\submitted{To be submitted to ApJ}
\title{Likelihood Functions for Galaxy Cluster Surveys}

\author{Gilbert P. Holder \altaffilmark{1,2}}
\altaffiltext{1}{CIAR Scholar}
\altaffiltext{2}{Canada Research Chair in cosmological astrophysics}
\affil{Department of Physics, McGill University, 3660 rue University, Montreal,
QC}
\email{holder@physics.mcgill.ca}

\begin{abstract}
Galaxy cluster surveys offer great promise for measuring cosmological
parameters, but survey analysis methods have not been widely studied. Using
methods developed decades ago for galaxy clustering studies, it is
shown that nearly exact likelihood functions can be written down for
galaxy cluster surveys. The sparse sampling of the density field by
galaxy clusters allows simplifications that are not possible for galaxy
surveys. An application to counts in cells is explicitly tested using
cluster catalogs from numerical simulations and it is found that 
the calculated probability distributions are very accurate at masses
above several times $10^{14}h^{-1} M_\odot$ at $z=0$ and lower masses
at higher redshift. 
\end{abstract}


\section{Introduction}

The idea of using cluster surveys as probes of dark energy has generated
considerable interest of late 
\citep{carlstrom02, haiman00, weller01,hu03,majumdar03,lima05}. 
As the largest virialized objects in the
universe, it is thought that clusters may be less affected by unknown
or poorly understood astrophysical mechanisms and it is hoped that it
will be reasonably straightforward to compare observed catalogs with
theoretical estimates.

While many recent works have investigated the power of upcoming surveys
and the challenges due to poorly understood selection effects, there has
been relatively less attention in the literature to methods of analysis.
It has been generally assumed that the problem can be separated into
analyzing the differential number counts as a function of redshift and
investigating the correlations (either angular or three-dimensional)
in the cluster catalog. Studies of galaxy surveys have traditionally
used either the power spectrum or $n$ point correlation functions to
construct likelihood estimates, and most studies of the galaxy cluster
abundance have assumed Poisson statistics. However, it has been recently
emphasized that Poisson statistics for galaxy cluster surveys may not
be a particularly good assumption \citep{hu03b,evrard02}.

The importance of sample variance considerations is two-fold. It broadens
the probability distribution of the counts in cells, weakening constraints
on the true number density, and the amount of this broadening is in
principle directly calculable from theory, adding new information which
in principle can be used to tighten constraints. As will be shown
below, the effect of sample variance on cluster counts is not
negligible and treating the counts as a Poisson process is 
incorrect.

In this work, we will go back to the work of \citet{white79}
in galaxy clustering and update it to the context of galaxy cluster
studies. In particular, we will derive the correct likelihood function
to use for binned estimates of the cluster number counts
and we write down an explicit expression for
the likelihood of observing a particular galaxy cluster catalog as a
function of theoretical models for the correlation structure of the
density field. We will present some very preliminary tests using the
catalogs from the Hubble volume simulation, demonstrating the regimes
where these methods are robust and where more work clearly needs to
be done. 

\section{Formalism}

We adopt the formalism of Dodelson, Hui, \& Jaffe (1997; hereafter
DHJ97), which in turn closely follows the development 
of White (1979; hereafter W79).  All that follows below is clearly 
outlined in W79 and the interested reader is referred to that paper
for a full exposition.

\nocite{white79,dodelson97}

The fundamental assumption is that the probability of observing an object 
at a particular location (in the absence of any other knowledge) is
simply proportional to the number density of objects:
\begin{eqnarray}
P(X_1)  =  n w_1(\vec{x_1}) dV_1  
\end{eqnarray}
where $n$ is the number density, $w_1(\vec{x_1})$ is the survey weight
at that point (the probability of detecting an object if it was at that
point), and $dV_1$ is an infinitesimal volume around the position 
$\vec{x_1}$.

By similar reasoning, the probability of observing two objects at two
points in space $\vec{x_1}$ and $\vec{x_2}$ can be expressed as
\begin{eqnarray}
P(X_1 X_2)  =  n^2 [ w_1(\vec{x_1}) w_1(\vec{x_2})+ 
	w_2(\vec{x_1},\vec{x_2})] dV_1  dV_2 
\end{eqnarray}
where $w_2(\vec{x_1},\vec{x_2})$ is the two-point correlation function,
indicating the excess probability (over random) of detecting two objects
at the given positions.

This reasoning can be extended to three objects:
\begin{eqnarray}
P(X_1 X_2 X_3) &  = & n^3 \bigl[ w_1(\vec{x_1}) w_1(\vec{x_2}) w_1(\vec{x_3}) +  \\
  & & \sum w_2(\vec{x_i},\vec{x_j})  w_1(\vec{x_k})  + \nonumber \\
& &  w_3(\vec{x_1},\vec{x_2},\vec{x_3})\bigr] dV_1  dV_2 dV_3 \nonumber
\end{eqnarray}
with $w_3$ the three-point correlation function, and the sum refers
to summing over all unique $i,j,k$. Obviously this can
continue to even longer lists of objects.

A given catalog will contain $N$ objects, and the probability of detecting
those $N$ objects at those positions can be constructed from the
hierarchy of correlation functions, but a very useful component of the
catalog is that {\em there are no objects detected at other points.}
A quantity of interest is therefore the probability of detecting $N$ objects
at observed $\vec{x_i}$ and no other objects at the other positions.

A recipe for incorporating this extra information was described in W79.
Instead of using the functions $w_i$ that were used above, one can
simply replace each occurrence of $w_i$ with $W_i$, where this latter
quantity is simply
\begin{eqnarray}
 W_i(\vec{x_1},\vec{x_2},...,\vec{x_i}) =  \hskip 2in \\
 \sum_{j=0}^{\infty} { (-n)^j \over j!}
\int ...\int w_{i+j}(\vec{x_1},...\vec{x_{i+j}} ) dV_{i+1}...dV_{i+j} \nonumber
\end{eqnarray}

It is nice to have an exact likelihood formulation for a given catalog, but
a hierarchy of infinite sums can be a bit unwieldy. DHJ97 noted that for
galaxy surveys there can easily be collections on the order of $10^{500}$
terms, even in the limit of Gaussianity and weak correlations. It will be
shown below that galaxy cluster surveys the number of
terms is quite manageable, at least in the limit of the cluster
distribution being Gaussian and for a reasonably sparse survey. Cluster
surveys can effectively be considered as perturbatively different from
a Poisson distribution, allowing the probability distributions to be
calculated quite efficiently.

In the Gaussian limit, it is useful to consider a few $W_i$:
\begin{eqnarray}
W_0(V) =  -nV + { n^2 \over 2} \int dV_1 dV_2 w_2(\vec{x_1},\vec{x_2}) 
\\
W_1(\vec{x_1}) =  1 -  n \int dV_2 w_2(\vec{x_1},\vec{x_2}) 
\\
W_2(\vec{x_1},\vec{x_2}) =  w_2(\vec{x_1},\vec{x_2}) 
\end{eqnarray}
where $V$ is the survey volume and the $w_1(\vec{x_i})$ have all
been set to unity for simplicity. Note that this is not a requirement
and is a straightforward way to include survey non-uniformity.

This allows an expression to be written down for observing a collection
of points as a function of a theory which predicts the number density
and two-point correlation function. To first order in the two-point
function (DHJ97),
\begin{eqnarray}
\label{eqn:catprob}
ln(P) & = & -nV + \\ \nonumber 
& & \hskip -0.3in ln\left[ \prod_a W_1(\vec{x_a}) + 
{1 \over 2} \sum_a \sum_{b\neq a} 
{W_2(\vec{x}_a,\vec{x}_b)\over W_1(\vec{x}_a)W_1(\vec{x}_b)}
\prod_c W_1(\vec{x}_c) \right] \nonumber
\end{eqnarray}

In the limit of sample variance being negligible (i.e.,
in the limit where $W_i$ approaches $w_i$), this approaches the
likelihood one would expect for the number counts (the first term
and the first term in brackets approaches the Poisson expression)
combined independently with the correlation function. The
second term in brackets closely resembles the likelihood one would obtain
for the correlation function assuming that the probability of observing
a pair at a given pair of locations is a Poisson process with mean given by
$n^2 (1+w_2)$. When sample variance is not negligible, however, the
probability will not be cleanly separable.

Furthermore, the same formalism was used by W79 to write down a
compact expression for the counts-in-cells probability:
\begin{eqnarray}
P[\Phi_N(V)] = { (-nV)^N \over N!} {d^N \over dn^n} exp[W_0(V)]
\label{eqn:p_phi}
\end{eqnarray}
where $\Phi_N(V)$ indicates observing $N$ objects (and only $N$) in a
volume $V$.

Again, assuming weak correlations and working in the Gaussian limit,
the counts in cells probability can be derived. Some details are derived
in the appendix, but the result is straightforward. The probability of
observing $N$ objects in a volume $V$ (to first order in the clustering)
is 
\begin{equation}
P[\Phi_N(V)] = {x^N e^{-x} \over N!}\bigl[1- n^2 V^2 {\bar{w} \over 2}\bigl( 
{ N-N^2 \over x^2} + 2{N \over x} -1 \bigr)\bigr ]
\label{eqn:countsprob}
\end{equation}
where $N$ is again the observed number, $x \equiv nV$, and $\bar{w} \equiv
1/V^2 \int dV_1 dV_2 w_2(\vec{x_1},\vec{x_2})$ is the mean correlation function
over the survey geometry. In calculations in this paper we use the second
order expression to generate the figures, but a careful analysis of terms
indicates that if the second order term is important then there is a good
chance that all higher order terms must be included.

Where might this expression be expected to break down? In W79 it is
clear that $W_0$ starts to become a challenging series when 
$n V \bar{w}$ becomes of order 0.1 or higher. We should therefore expect
that this likelihood calculation will be most effective in that case
and that it is suspect outside that regime. The breakdown appears to
arise because the assumption of Gaussianity must always break down:
the density field cannot be negative. With sparse sampling the probability
of encountering a negative density is negligible and the assumption
of Gaussianity is sufficient. However, at higher number densities the
intrinsic non-Gaussianity becomes important.\footnote{As this work
neared completion a preprint from Hu \& Cohn appeared where
they correct for this by ignoring those excursions of the density field 
and treating the remaining field as Gaussian.}

Notice that this approach is straightforward to extend to objects that
are tagged with an observable such as flux or richness. 
Going back to eqn \ref{eqn:catprob}, it should be sufficient to replace
the initial number density with a sum over the individual number densities
in each observable bin (e.g., theoretical number density as a function 
of optical richness or SZ flux), and then keep in mind that each
$W_i$ will depend on the mass of the object or pairs of object
under consideration. The efficacy of this technique is
currently a topic of investigation.

\section{Counts in Cells}

We use the cluster catalogs derived from the Hubble volume simulations 
\citep{colberg00}\footnote{http://www.mpa-garching.mpg.de/Virgo/hubble.html}
to test the range of validity of
the likelihood expression derived for the counts in cells. A straightforward
test is to use the $z=0$ cluster catalogs, where evolution effects
can be neglected. We further specialize to the $\tau CDM$ simulation,
where the transfer function is easy to reproduce. The simulation 
parameters are $\Omega_m=1.0, \sigma_8=0.6$ and $\Gamma=0.21$ and a
box size of $2000 h^{-1} Gpc$.

The volume is divided into subregions of cubic geometry and the
above expressions are tested for a variety of cell sizes and mass
thresholds. The number of clusters within each cell is then used to 
construct the probability distribution of the counts in cells.
Mass thresholds were used to construct counts in cells for all objects
above a threshold mass, with the mass defined as the number of
particles (each of mass $2.2 \times 10^{12} h^{-1}M_\odot$) found by
the friends of friends algorithm with a linking parameter of 0.2.

The mean correlation function of the matter distribution  was derived as 
\begin{eqnarray}
\bar{w} & = & {1 \over V^2} \int dV_1 dV_2 w_2(\vec{x_1},\vec{x_2}) \\\nonumber 
& = & {1 \over 8 \pi^3 V^2} \int d^3k W(\vec{k})W^*(\vec{k}) P(k) 
\end{eqnarray}
where $W(\vec{k})$ is the window function for the cells (not related
to the $W_i$ above!). For cubic cells, the window function is
simply a product of $sinc$ functions.

The mean correlation function of the galaxy clusters needs to be
corrected for the bias function. This is non-trivial; several
fitting functions exist in the literature but typical errors 
appear to be on the order of 10\% when applied in a regime where
they have not been calibrated \citep{colberg00}.  We use the bias relation of
\citet{sheth01} with the mean bias for the cluster sample
above a mass threshold derived using a mass function weighted
bias factor. The mass function of \citet{jenkins01} was used,
which has been calibrated using the simulation used here. In principle,
given a mass correlation function one can estimate the bias as a function
of an observable cluster property and use this to estimate
the cluster mass as a function of cluster properties. However, current
uncertainty in the theoretical understanding of bias as a function of mass
in arbitrary cosmologies complicates any such undertaking.

An example is shown in figure 1. Taking a relatively high 
threshold of $6.6 \times 10^{14} h^{-1} M_\odot$ and a
cell size of 154 $h^{-1}$Mpc, the counts in cells distribution
can be synthesized. We average over ten random cell offsets to
smooth the observed distribution and compare it to the exact 
calculation of eqn \ref{eqn:countsprob}. 
In the limit of massive (and therefore rare and sparse)
clusters the likelihood for the counts in cells is quite accurate.

\begin{figure}
\plotone{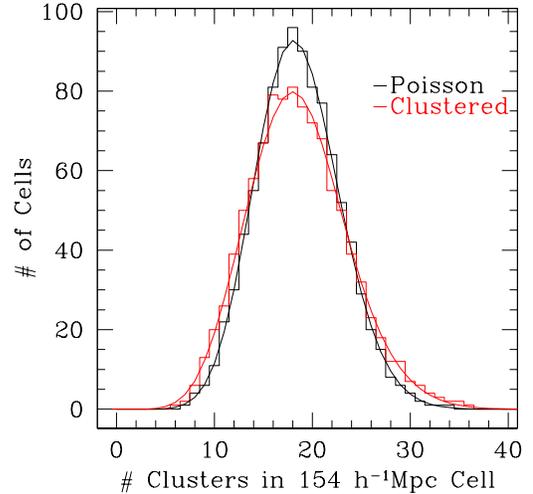}
\caption{Distribution of cell counts for clusters above
$6.6 \times 10^{14} H^{-1}M_\odot$ in Hubble Volume simulation
compared to prediction from theory (broader distribution with
lower peak)
and corresponding result from same clusters with randomized positions
(narrower distribution with higher peak).}
\end{figure}

The assumption of a Gaussian random field will break down when
$nV\bar{w}$ is of order $10^{-1}$ (W79). To investigate the nature
of the breakdown, we show in figure 2 the first breakdown for
cells of size 100 Mpc and 133 Mpc. At higher masses (i.e., lower
number densities) the theoretical result matches the simulations
quite well. It is clear that there is noticeable deviation starting
around $nV\bar{w}\ga 0.5$.
The mass thresholds were
selected to straddle the transition. At yet lower masses, the
theoretical result breaks down fairly comprehensively, developing
bimodality. This is an indication that the theory is simply not well
defined in this regime. It is impossible to set up a distribution
of positive mass concentrations with vanishing correlation functions
above the two-point function.

\begin{figure*}
\plottwo{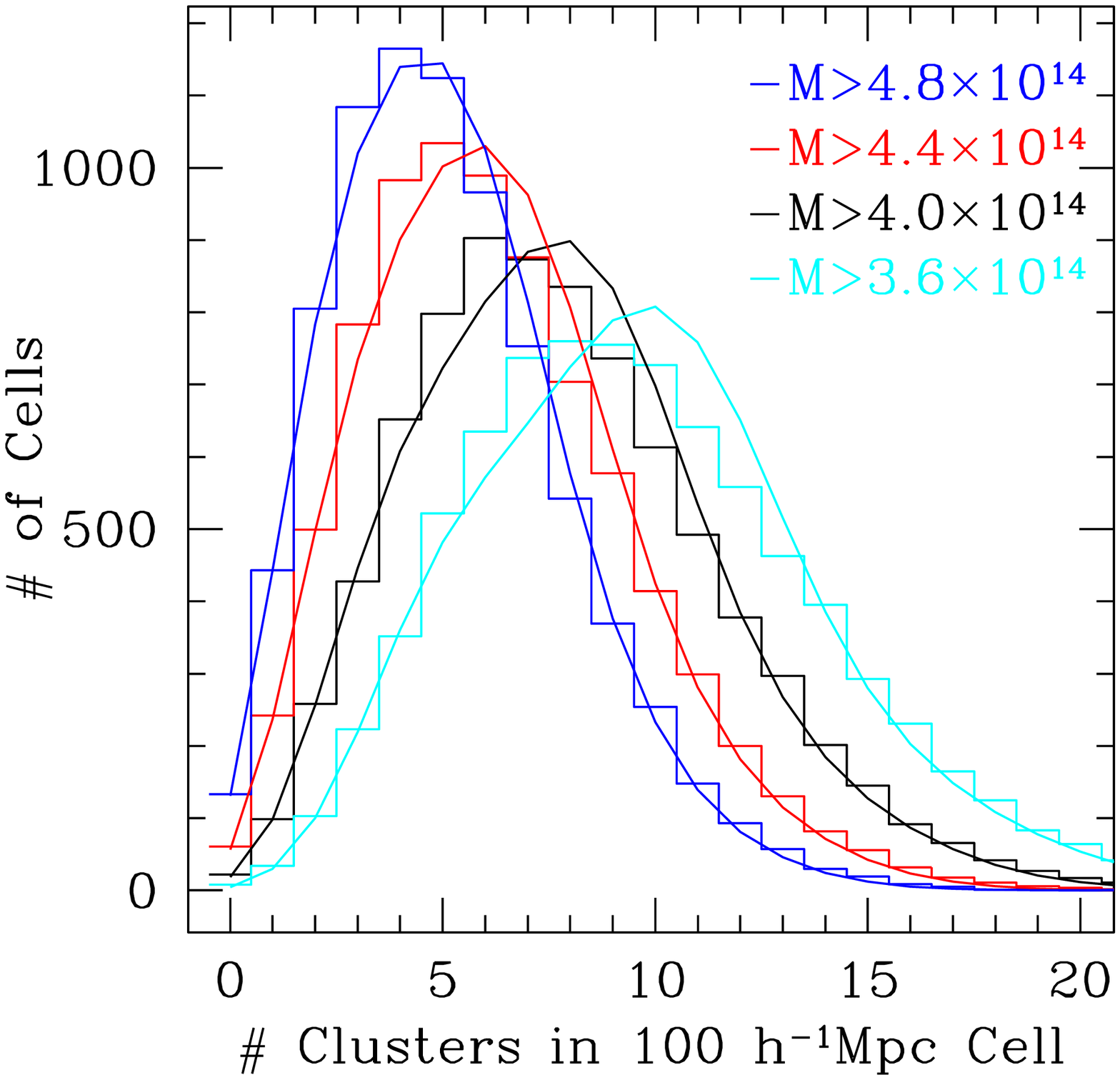}{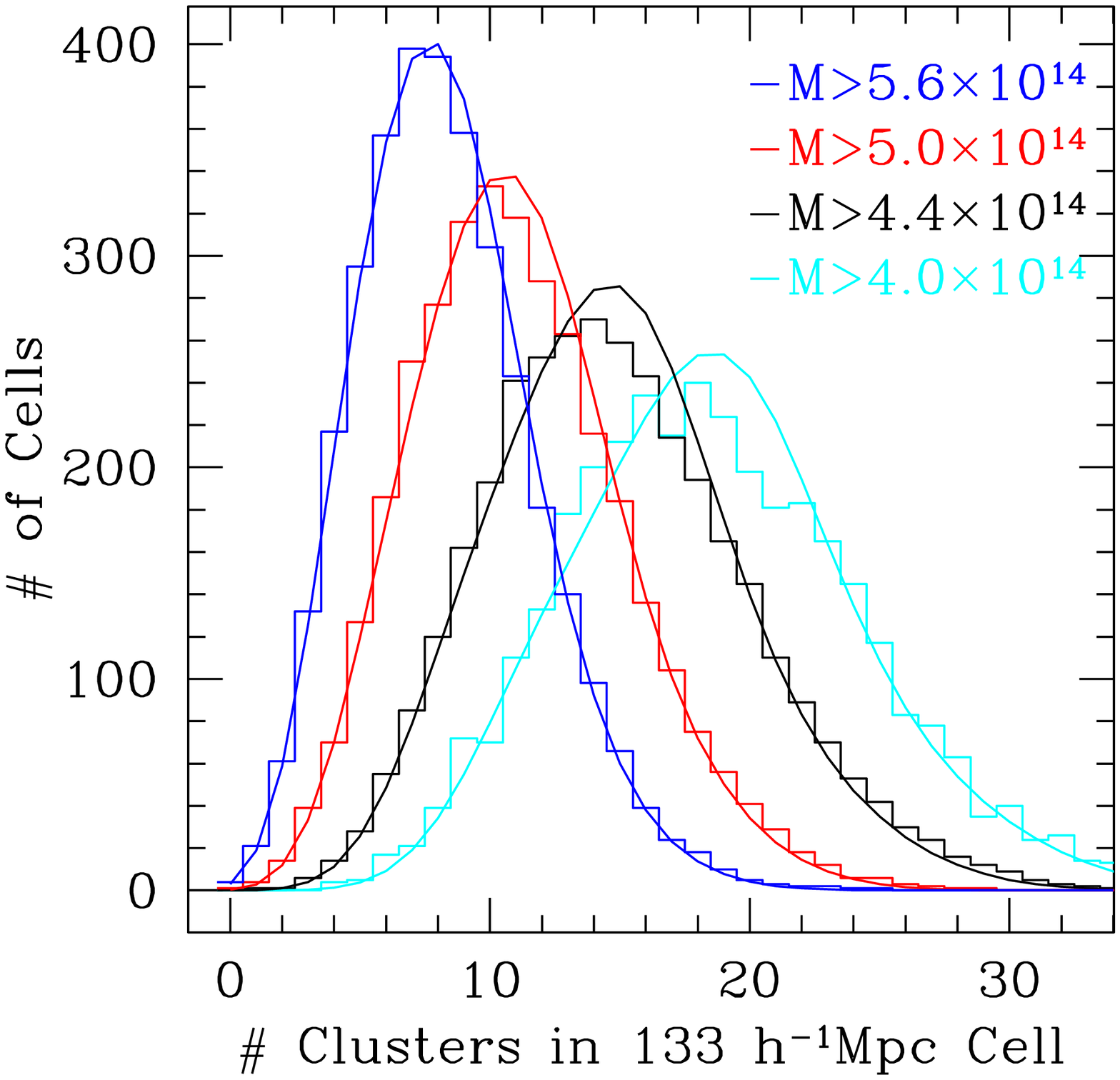}
\caption{Number of cells in simulation volume with given number of galaxy
clusters above the mass threshold. The mass threshold is varied
for two values of fixed cell
size: 100 $h^{-1}$Mpc (left) and 133  $h^{-1}$Mpc (right). In the left
panel, the values of $nV\bar{w}$ range from 0.44 to 0.68 (left to right in
the figure) and in the right
panel they range from 0.34 to 0.6 }
\end{figure*}

As the mass threshold is lowered, clusters become more numerous.
As the cell size decreases, the mean number per cell will decrease
but the mean correlation function over the cell volume will increase.
For reasonable power spectra it turns out that the number per cell
is slightly more important in this case for the purposes of 
calculating $nV\bar{w}$, making it slightly advantageous to move to
smaller cell sizes.

To investigate the regime where this likelihood approach can be
trusted, a grid of masses and cell sizes was investigated. The
quantity $nV\bar{w}$ was calculated and the lines in figure 3
indicate the locus of pairs of $M$ and cell size that lead to
$nV\bar{w}=0.5$. For a $\Lambda CDM$ model with $\sigma_8=0.9$
the likelihood approximation breaks down below $4 \times
10^{14} h^{-1}M_\odot$, with the $\tau CDM$ breaking down around
the same mass. However, this is a very steep function of the
amplitude of the power spectrum. For example, using $\sigma_8=0.81$
drops the threshold of applicability to below
$3 \times 10^{14} h^{-1}M_\odot$. This is particularly relevant
because this corresponds to the amplitude of fluctuations at
$z=0.2$ if $\sigma_8=0.9$. 
Any survey for galaxy clusters at $z \ga 0.2$ and
targeting masses above a few $\times 10^{14} h^{-1}M_\odot$ is
squarely is the regime where this likelihood calculation can
be used. 

\begin{figure}
\plotone{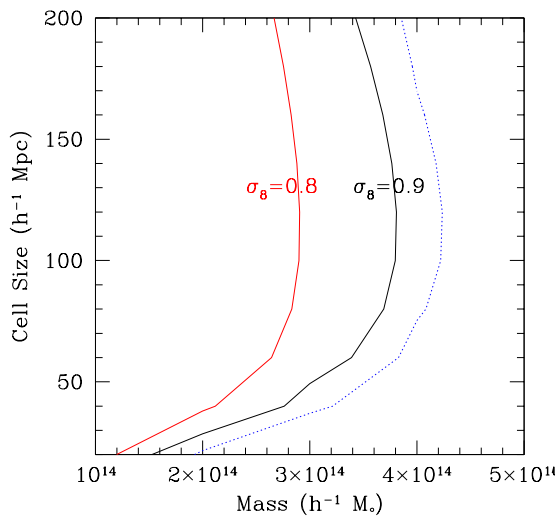}
\caption{Combinations of mass and cell size that lead to $nV\bar{w}=0.5$
assuming $\Omega_M=0.3, \Omega_\Lambda=0.7$, and power spectrum
shape parameter $\Gamma=0.21$ for two values of $\sigma_8$. The
dotted line corresponds to the $\tau CDM$ model that was used in the
Hubble Volume simulations. The parameter space to the right of each line
is in the regime of being easy to implement the likelihood formulation
in this paper.}
\end{figure}

\section{Outlook}

The preliminary (and very highly idealized) tests outlined above 
demonstrate that many cluster surveys can be considered to be in the
regime where catalog likelihood methods will be readily applicable.
Tests on counts in cells at $z=0$ indicate that the non-Gaussianity
of the cluster distribution becomes important around masses of several
times $10^{14}h^{-1}M_\odot$. Applications of the cluster catalog
likelihood (eqn \ref{eqn:catprob}) will likely have similar problems
at lower masses and may also have difficulties at small separations.
This is a topic of ongoing investigation.

The tests performed in this work used a $z=0$ simulation snapshot,
rather than a lightcone selected volume. This was done to cleanly
separate the possible complications and there is no conceptual
problem with generalizing this to include redshift evolution. 
In fact, to generalize this one simply goes back to 
eqn \ref{eqn:countsprob} and allows $n,V,\bar{w}$  to be functions of 
$z$ and cosmological 
parameters, including the evolution of the relevant window functions.
Non-cubic cell volumes that are functions of cosmological parameters
and redshift will certainly complicate the analysis, but all the
results presented here should be directly applicable.

There are several directions for more work in this area. Tests on
existing cluster catalogs will yield insight into exactly how hard
it is to implement these ideas in practice. One of the lessons learned
is that this method works best on sparse catalogs (i.e., massive
clusters), so in existing data it would be most easily tested on
the most massive X-ray cluster surveys. It is unfortunate that the
method appears to break down in a regime where there are high quality
local cluster surveys. 

On the theoretical side,
an obvious next step is to apply these methods using clusters that
have been additionally tagged with an observable such as X-ray or SZ flux
or optical richness. Within this formalism such a modification is
entirely straightforward, as all that is required are number densities
and correlation strengths. Incorporating scatter in something like
a mass-observable relation would be in principle straightforward, as
would allowing for non-trivial (non power-law) evolution in such 
relations. However, uncertainty in the bias relations as a function
of cosmology and mass must be improved to realize the full potential
of these methods.

Given that the breakdown in this formalism appears to coincide with
the onset of non-linearity, it may be possible to extend these
results to lower masses with some modelling of the effects of
weak non-linearity. The excellent match between theory and simulations
at higher masses demonstrates that non-linearity and non-Gaussianity
are not important in the high-mass regime.

\section{Summary and Conclusions}

In this paper we outlined a likelihood approach to galaxy cluster
surveys that is uniquely applicable to the case of massive galaxy
clusters. Essentially, for small enough sample and/or weak enough
clustering strengths one can treat cluster catalogs as being
expected to be drawn from a distribution that can be thought of
as a Poisson distribution with some perturbations.
By virtue of being only perturbatively different from a Poisson 
process, cluster catalogs lend themselves well to efficient
likelihood estimation techniques that naturally include both shot noise
and sample variance. This does not require an arbitrary conceptual
breaking of the problem into ``counts'' and ``clustering'' but instead
simply addresses the question of how likely it is to obtain a set of
points given an assumed underlying correlation structure.

For galaxies, the number of terms involved in the estimate is
prohibitive, and it is also important to understand the higher
order correlations. That does not appear to be the case for 
massive galaxy clusters. The Gaussian weak correlation limit does a
very good job of matching the counts in cells distribution for
massive clusters in the Hubble volume simulation. As the mass threshold
is lower, and the number density therefore increases, some disagreement
arises, likely due to the number density becoming high enough that
the distribution becomes sensitive to non-Gaussianity. 

It should be emphasized that it is clearly
not an advantage to have fewer clusters. One could always thin the 
number density of a sample until the catalog is in a regime where a
catalog likelihood can be calculated. This would obviously be throwing
away a huge amount of information and it seems unlikely that the gain
in simplicity would be worth the loss of information.

These techniques are straightforward to generalize to explicitly include
inhomogeneous survey selection and differential number densities and
clustering strengths as a function of galaxy cluster properties. 

With galaxy cluster surveys now being used for cosmological parameter
estimation, it is essential to include the effects of sample variance
to obtain accurate parameter estimates, so obtaining accurate
probability distributions for counts in cells should be a high priority.
The techniques here are a step in that direction, showing that such
accurate distributions are easily calculated for massive galaxy clusters.

As this article neared completion, an article appeared by Hu \& Cohn
discussing very similar ideas from a slightly different perspective.
The results here largely agree with theirs in the aspects where they
overlap. \nocite{hu06}

\acknowledgements{I have had useful discussions with Wayne Hu
and have benefitted from an NSERC Discovery
grant, the Canada Research Chairs program, and the Canadian Institute
for Advanced Research.}

\appendix
\section{Details of the Counts in Cells Likelihood}

The calculation of the likelihood for counts in cells can be derived
from equation \ref{eqn:p_phi} with straightforward algebra and accounting in 
the limit of weak clustering. Some of the details are spelled out below.

\begin{eqnarray}
P[\Phi_N(V)] = 
{ (-nV)^N \over N!} {d^N \over dn^n} exp[W_0(V)]
\end{eqnarray}

Define $x\equiv nV$ and using $W_0 = -x + x^2 \bar{w_2}/2$, the first
step is to expand the exponential term as a Taylor series. For convenience, 
define $\alpha \equiv \bar{w_2}/2$ and the counts in cells probability
distribution can be written

\begin{equation}
P[\Phi_N(V)] = 
{(-x)^N \over N!} {d^N \over dx^N}\left( \sum_{i=0}^{\infty} {1 \over i!}
(-x)^i [1- \alpha x]^i \right)
\end{equation}

Expanding the part inside the sum with the binomial expansion gives
\begin{equation}
P[\Phi_N(V)] = 
{(-x)^N \over N!} {d^N \over dx^N}\left( 
\sum_{i=0}^{\infty} 
\sum_{k=0}^{i} 
{1 \over k! (i-k)! }
(-x)^{i+k} \alpha ^k \right)
\end{equation}

The effect of the derivatives is that all terms in the sums with
$i+k<N$ are set to zero and all other terms are multiplied by
$x^{-N}(i+k)!/(i+k-N)!$:
\begin{equation}
P[\Phi_N(V)] = 
{(-x)^N \over N!} \left( \sum_{i=0}^{\infty} \sum_{k=max(0,N-i)}^{i} 
(-1)^N {(i+k)! \over k! (i-k)! (i+k-N)!} \alpha ^k (-x)^{i+k-N} 
\right)
\end{equation}

Up to this point, the only assumption implicit in this analysis is
that the correlation functions beyond the two-point function are
negligible. To see the structure in the probability distribution, 
we now assume that $\alpha$ is small and investigate the relevant
terms in the above equation. In particular, for small $\alpha$
the only important terms will be terms with small $k$. This 
immediately identifies the important terms as being the ones
with $i$ close to $N$. 

For the terms with $k=0$, we are left with the sum
\begin{equation}
k=0: \ \ \sum_{i=N}^{\infty} { 1 \over (i-N)!} (-x)^{i-N} = e^{-x}
\end{equation}
which reproduces the Poisson probability in the case of zero 
correlations.

To first order in $\alpha$, one can move to one lower value of $i$:
\begin{equation}
k=1: \ \ \alpha \sum_{i=N-1}^{\infty} {(i+1)! \over (i+1-N)!(i-1)!} (-x)^{i+1-N} = 
-\alpha e^{-x}(N^2-N-2xN+x^2)
\end{equation}
and to second order in $\alpha$ we can continue down: 
\begin{eqnarray}
k=2: \ \ \alpha^2 \sum_{i=N-2}^{\infty} {(i+2)! \over 2(i+2-N)!(i-2)!} 
(-x)^{i+2-N}  & = &  \nonumber 
{\alpha^2 \over 2} e^{-x} (-8xN+12xN^2+6x^2N^2-6x^2N- \\
& & 4x^3N+11N^2-6N^3-6N-4xN^3+x^4+N^4 )
\label{eqn:secondorder}
\end{eqnarray}
with subsequent orders obtained similarly. These sums are straightforward to 
implement using software such as Maple but can become tedious. In all that
was implemented in this work the series was truncated at second order.

The sums have an interesting property that each term seems to be of order
$(\alpha x^2)^k$, since $N$ is of the same order as $x$. At each level
the terms are actually of lower order in $x$ due to cancellations in the
sums. For example, for $x=N$ the expression in brackets in equation 
\ref{eqn:secondorder} reduces to $3N^2-6N$, making this term roughly
$\alpha^2 x^2$. Nonetheless, it is clear that simply having small
$\alpha$ is not sufficient. For this formalism to be useful it is
required that $\alpha x$ be small, at least, and it is preferable
that $\alpha x^2$ be small to be sure that all higher order terms are
negligible. It is apparent from the expression for $W_0$ that bad things
are happening as $\alpha x$ is becoming large, since it is becoming
dubious that the void probability function is bounded to be less than
one (i.e., $W_0$ must be negative). This is likely a breakdown of the physical
picture of Poisson sampling of a Gaussian random field, rather than a
breakdown of any approximations. It is impossible to Poisson sample with
a negative mean, which a Gaussian random field allows. The requirement
that the density is non-negative necessitates some amount of non-Gaussianity.
With enough samples, this non-Gaussianity must manifest itself in the
observed distribution.

%\bibliography{/homes/tapajo/holder/Tex/composite}

\end{document}